# The Effect of Choice of Metric and Scan Length on Reliability in Resting-State fMRI

**Yu Huang[1,2,*] , Philip T. Reiss[3], Seonjoo Lee[2,4,5], R. Todd Ogden[2,4]**

[1] Department of Population Health, New York University Grossman School of Medicine, New York, New York, USA

[2] Department of Biostatistics, Mailman School of Public Health, Columbia University, New York, New York, USA

[3] Department of Statistics, University of Haifa, Haifa, Israel

[4] Department of Psychiatry, Columbia University, New York, New York, USA

[5] Mental Health Data Science, New York State Psychiatric Institute, New York, New York, USA

**Abstract**

Resting-state fMRI (rs-fMRI) is widely used to investigate brain functional connectivity, but the reliability of these measurements remains a key concern for ensuring reproducibility. The distance-based intraclass correlation coefficient (dbICC) generalizes classical ICC to more general data types, making it well-suited for assessing the reliability of measures of functional connectivity. In this study, we applied dbICC to assess the reliability of rs-fMRI data from the Midnight Scanning Club (MSC) dataset, which consists of 10 subjects, each undergoing 10 sessions of 30-minute rs-fMRI scans. The functional connectivity was estimated using Pearson's correlation coefficients between all pairs of brain regions, resulting in a correlation matrix for each session. We compared two distance metrics—the widely used Frobenius metric and the Affine Invariant Riemannian Metric (AIRM) selected to respect the geometry of the space of covariance matrices—to evaluate how the choice of metric affects the reliability of estimating correlation. In addition, we investigated the impact of scan length and time interval between sessions on reliability. Results based on each metric agreed in some respects but disagreed in others, illustrating the impact of choice of metric. We also found that longer scan lengths significantly improve reliability, while the time interval between sessions has less impact.

## 1. Introduction

Resting-state functional magnetic resonance imaging (rs-fMRI) has become an essential tool for investigating brain functional connectivity. It has been widely used to identify resting-state networks (Van Den Heuvel & Hulshoff Pol, 2010) and to provide diagnostic and prognostic information for neurological and psychiatric disorders (Lee et al., 2013). However, a critical challenge in rs-fMRI research lies in ensuring the reliability of connectivity measurements, which is fundamental for reproducibility across studies and applications.

[*] Corresponding author: Yu Huang (yh5311@nyu.edu)

The intra-class correlation coefficient (ICC) is widely used for reliability assessment in many applications and is defined as the ratio of between-subject variance to total (within- plus between-subject) variance (Shrout & Fleiss, 1979). It has been applied in a wide range of contexts including physical measurements, psychological testing, and educational scoring. ICC has also been used to evaluate the reliability of fMRI results. For instance, Caceres et al. (2009) used voxel-wise ICC maps to characterize reliability at the voxel level and summarized ROI reliability by taking the median of voxel-wise ICC values within a region to assess the consistency of voxel-wise activations across repeated fMRI sessions in task activation studies. Similarly, Shehzad et al. (2009) examined the test–retest reliability of resting-state functional connectivity by computing ICC for each pairwise functional connection, both at the voxel level and after aggregation to ROI-level, demonstrating that functional connectivity measures show moderate to high reliability across sessions.

When the observed data consist of scalar values, ICC is straightforward to compute since the variance components are also scalars. However, when observations are vector-valued, their variance must be expressed as a covariance matrix. Thus, the extension of ICC to vector-valued data is not immediately obvious, since the ratio of covariance matrices is not well defined. A simple extension is to vectorize each observation and compute scalar ICC element-wise, as in Caceres et al. (2009). However, this approach neglects the inherent structure of covariance matrices. To address this limitation, the image intraclass correlation coefficient (I2C2) (Shou et al., 2013) extends classical ICC by vectorizing the entire image or connectivity matrix and defining reliability based on the covariance structure of the repeated measurements. Specifically, I2C2 decomposes the total variability into a between-subject component (true signal) and a measurement error component, and summarizes their magnitudes via the trace, yielding a scalar-valued reliability index. While ICC and I2C2 have been widely used for reliability assessment, such measures do not fully capture the geometric properties of high-dimensional fMRI data.

More recently, the distance-based intraclass correlation coefficient (dbICC) has been introduced by generalizing ICC in terms of squared distances between pairs of observations. This allows application to more general data types, such as curves, graphs, shapes, and networks provided that there is a distance metric that is appropriate for the data objects (Xu et al., 2021). The choice of distance metric is critical when applying dbICC. The Frobenius metric, which is equivalent to the Euclidean metric when the matrices are vectorized, is commonly used in fMRI reliability measurement due to its simplicity and broad applicability. This representation is conceptually related to the framework used in I2C2, which treats images as high-dimensional vectors. However, the Frobenius metric was originally designed for vectors in Euclidean space. While a correlation matrix can be represented as a vector, such an approach does not fully capture the geometric structure of functional connectivity (You & Park, 2021).

Correlation matrices, such as those representing functional connectivity, are symmetric and positive definite (SPD). The set of SPD matrices of a given size forms a cone-shaped Riemannian manifold. Directly applying the Frobenius metric to these matrices may fail to account for their underlying geometry. To address this issue, Riemannian metrics have been proposed as an alternative (You & Park, 2021), which takes into account the corresponding geometric structure. The effectiveness of a Riemannian metric has been demonstrated in previous work. Using EEG motor imagery data, Barachant et al. (2012) computed covariance matrices for each trial, derived both class-specific Riemannian means and Euclidean means, and classified trials by their distances to these means. They found that Euclidean distances failed to separate the right-hand and left-hand motor imagery classes, whereas the Riemannian distance yielded clear class separation and higher classification accuracy. This empirical example shows that the Riemannian metric can improve the analysis of SPD matrices relative to the Frobenius metric. The affine-invariant Riemannian metric (AIRM) is commonly used on the space of SPD matrices (Pennec et al., 2006), and in our study we adopt this metric as well.

A meta-analysis of 25 studies reported relatively low reliability for edge-level functional connectivity, defined as the correlation between the time series of a pair of ROIs, with a mean conventional scalar ICC of 0.29 (Noble et al., 2019). These studies focused on fMRI in the human brain, including both resting-state and task-based data from healthy controls and clinical populations, and applied various ICC models, including one-way random, two-way random, and two-way mixed effects models with both single-measure and average-measure forms (Shrout & Fleiss, 1979). The low ICC values raise concerns about measurement consistency in fMRI research (Noble et al., 2019). One approach to improving reliability is simply to increase the amount of data collected—in the case of rs-fMRI this could be accomplished by extending scan length (Birn et al., 2013). However, longer scans are not always feasible in practice. Subjects may become uncomfortable when staying in the scanner for extended periods, and they may have difficulty staying still, leading to increased motion artifacts and degraded data quality. Moreover, existing studies have found that the improvement in reliability shows rapidly diminishing returns as scan length increases (Mueller et al., 2015; Tomasi et al., 2017; Van Dijk et al., 2010). Thus, researchers have to weigh the trade-offs among collecting more data, maintaining subject compliance, and image quality. Identifying an optimal scan length that balances these factors is therefore important.

In addition to scan length, the time interval between repeated scans may also influence reliability (Noble et al., 2019). This is particularly relevant for longitudinal studies, such as those in which measurements are taken before and after interventions. Many clinical interventions take place over several days or even weeks, and it remains unclear whether measurement of functional connectivity remains stable over such extended intervals.

To investigate these questions, we analyze the Midnight Scanning Club (MSC) dataset (Gordon et al., 2017), which includes data from 10 individuals, each of whom underwent rs-fMRI scans on 10 consecutive days, with each session lasting 30 minutes. This design allows us to examine how reliability varies as a function of scan length. In addition, we assess whether the time interval between scans influences reliability by comparing dbICC estimates from sessions separated by varying time gaps (e.g., on consecutive days vs. four days apart). These investigations can provide guidance for experimental design in fMRI studies, particularly in optimizing data collection strategies to enhance measurement reliability.

## 2. Methods

### 2.1 Dataset and Participants

The MSC dataset consists of 10 healthy, right-handed participants (5 females, age 24–34 years). Each participant completed 10 resting-state fMRI sessions over 10 consecutive days, with each session lasting 30 minutes, resulting in 818 time points (TR = 2.2 s) per session. The data was obtained from the OpenNeuro database and its accession number is ds000224 (https://openneuro.org/datasets/ds000224) (Gordon et al., 2017). During resting-state scans, participants kept their eyes open and fixated on a white crosshair presented against a black background. All data were acquired using a gradient-echo EPI sequence (TE = 27 ms, flip angle = 90°, voxel size = 4 mm × 4 mm × 4 mm, 36 slices). Further details of the image acquisition have been previously reported (Gordon et al., 2017). We used the preprocessed version of MSC dataset, which is available on OpenNeuro (https://openneuro.org/datasets/ds000224) (Gordon et al., 2017).

### 2.2 Functional Connectivity and Networks

The voxel-wise time-series were parcellated into 360 cortical regions of interest (ROIs) using the multi-modal parcellation atlas developed by Glasser et al. (2016). ROI-level time series were derived by averaging the BOLD signal across voxels within each of the 360 ROIs. For each session, we computed a 360 × 360 functional connectivity matrix by calculating Pearson correlations between the time series of all pairs of cortical regions. We refer to this as whole-brain functional connectivity.

In addition to the whole-brain functional connectivity, we also examined specific functional networks: the default mode network (DMN), frontoparietal network (FPN), cingulo-opercular network (CON), visual network (VIS), and somatomotor network (SMN) (Ji et al., 2019).

### 2.3 Distance-based ICC

We applied the distance-based intraclass correlation coefficient (dbICC; Xu et al., 2021) to assess the reliability of rs-fMRI data. The dbICC generalizes the classical ICC by redefining it in terms of arbitrary distances among observations. To define the dbICC we first review the ICC setup for scalar-valued data, in which $I$ individuals indexed by $i = 1, \dots, I$, each have a "true" characteristic or aspect $T_i$ drawn from a population with variance $\sigma_T^2$ and assumed to be independent across individuals. Each individual $i$ is measured $J_i$ times, yielding observations $X_{ij}$ for $j = 1, \dots, J_i$. Each observed value $X_{ij}$ is modeled as

$$X_{ij} = T_i + \varepsilon_{ij} \tag{1}$$

where the error terms $\varepsilon_{ij}$ are mutually independent, with mean 0 and variance $\sigma_\varepsilon^2$, and independent of the $T_i$'s. For distinct $j_1, j_2 \in \{1, \dots, J_i\}$, the ICC is the correlation between the $j_1$th and $j_2$th observations for individual $i$. Under the model (1), this correlation can be shown to equal the proportion of the total variance that is attributable to the true variance:

$$\rho = corr(X_{i1}, X_{i2}) = \frac{\sigma_T^2}{\sigma_T^2 + \sigma_\varepsilon^2}$$

We can rederive the ICC in terms of squared distances between observations. If we define the mean squared differences of measurements between individuals and within individuals as $\mathrm{MSD}_b = E_{i_1 \neq i_2}\left[\left(X_{i_1 j_1} - X_{i_2 j_2}\right)^2\right] = 2\sigma_\mathrm{T}^2 + 2\sigma_\varepsilon^2$ and $\mathrm{MSD}_w = E_{j_1 \neq j_2}\left[\left(X_{i j_1} - X_{i j_2}\right)^2\right] = 2\sigma_\varepsilon^2$, respectively, then the ICC can be expressed as

$$\rho = 1 - \frac{MSD_w}{MSD_b}. \tag{2}$$

The dbICC may be defined by redefining $MSD_b$ and $MSD_w$ in terms of distances $MSD_b = E_{i_1 \neq i_2}\left[d\left(X_{i_1 j_1}, X_{i_2 j_2}\right)^2\right]$ and $\mathrm{MSD}_w = E_{j_1 \neq j_2}\left[d\left(X_{i j_1}, X_{i j_2}\right)^2\right]$. Here, $d(\cdot,\cdot)$ can be any arbitrary distance between observations. As with ordinary ICC, higher dbICC values indicate higher reliability. Confidence intervals for dbICC were obtained using bootstrapping with bias correction (Xu et al., 2021).

### 2.4 Distance Metrics

In our study, we used both Frobenius metric and AIRM to measure the distances between brain functional connectivity matrices computed for each scan session. Let $X$ and $Y$ be two such matrices, with elements $X_{pq}$ and $Y_{pq}$ where $p = 1, \ldots, m$ and $q = 1, \ldots, m$. The Frobenius distance between $X$ and $Y$ was computed as:

$$d_F(X, Y) = \sqrt{\sum_{p=1}^{m} \sum_{q=1}^{m} \left(X_{pq} - Y_{pq}\right)^2}.$$

This distance is mathematically equivalent to the Euclidean distance between $\mathrm{vec}(X)$ and $\mathrm{vec}(Y)$, where $\mathrm{vec}(\cdot)$ denotes the vectorization operator, which converts a matrix (here $m \times m$) to a vector (here in $\mathbb{R}^{m^2}$).

The AIRM distance between $X$ and $Y$ (Pennec et al., 2006) is given by

$$d_R(X, Y) = \| \log\left(X^{-1/2} Y X^{-1/2}\right) \|_F,$$

where $\|\cdot\|_F$ is the Frobenius norm and log refers to the matrix log (Higham, 2008). Since these calculations require the input matrices to be SPD, we applied a regularization procedure to a few cases in which the correlation matrix was singular by adding a small value to the diagonal of the covariance matrix, followed by conversion to a correlation matrix for distance computation. We used the CovTools package (v0.5.4; Lee & You, 2021) to calculate distance matrices. All data analyses were performed using RStudio Version 2024.04.2+764 and R Version 4.4.1. The resulting distance matrices were also visually inspected to identify potentially anomalous sessions characterized by unusually large within-subject distances.

### 2.5 The Impact of Scan Length and Time Interval on Reliability

To examine how scan length affects reliability, we truncated the full BOLD time series to varying scan lengths. We used 100 scan lengths approximately evenly spaced between 221 and 818 time points. The lower limit of 221 was chosen to avoid unstable estimates associated with very short segments. After truncating the time series to the given length, we computed the corresponding connectivity matrices and the resulting distance matrices and dbICC values.

In the paper introducing dbICC, Xu et al. (2021) derived a generalized version of the Spearman-Brown formula, a classical result that shows how reliability increases when more repeated measures are taken. According to this generalized Spearman-Brown formula, changes in reliability with increasing scan length can be examined by plotting

$$\left[log(m), log\left(\frac{\hat{\rho}_m}{1-\hat{\rho}_m}\right)\right],$$

where $\hat{\rho}_m$ is the estimated dbICC corresponding to scan length $m$. The slope of the log–log plot can be interpreted as an empirical estimate of the power-law exponent governing how quickly $MSD_w$ decreases with scan length (Xu et al., 2021).

To evaluate the effect of time interval between sessions on reliability, we developed a time-interval-dependent version of the dbICC, in which the constant $MSD_w$ is replaced by mean squared within-subject distance as a function of time interval. This function can be estimated by fitting the nonparametric mixed-effects model

$$d_{i,j_1j_2}^2 = f\left(\Delta t_{i,j_1j_2}\right) + u_i + \varepsilon_{i,j_1j_2} \tag{3}$$

where $d_{i,j_1j_2}^2$ is the squared distance between sessions $j_1$ and $j_2$ from the same subject $i$, $\Delta t_{i,j_1j_2}$ is the time interval between those sessions, $f(\cdot)$ is a smooth function representing $MSD_w$ as a function of time interval, and $u_i$ is a subject-specific random intercept to account for repeated measures within subjects. We estimated $f$ using penalized splines (Wood, 2017), excluding pairs with time intervals of eight or nine days due to limited sample size.

Based on the function estimate $\hat{f}$ from the fitted model, we define a time-dependent estimate of dbICC as

$$dbICC(\Delta t) = 1 - \frac{\hat{f}(\Delta t)}{MSD_b}.$$

The test procedure of Wood (2013) can be used to assess whether mean squared within-subject distance is dependent on time interval, with the null hypothesis being that it is not (i.e., $f$ is a constant function).

## 3. Results

### 3.1 Frobenius distance versus AIRM

One participant (MSC08) was excluded from all analyses due to prolonged eye closures and excessive head motion (Gordon et al., 2017). Each of the remaining 9 participants had 10 scan sessions, for a total of 90 scans. We calculated all pairwise distances resulting in a 90 × 90 distance matrix for each of the two metrics considered: Frobenius metric and AIRM. **Figure 1** shows the distance matrices for all ROIs, DMN, and SMN, with darker colors representing larger distances. The FPN, CON, and VIS showed similar patterns as the DMN and are included in **Supplementary Figure 1**.

These figures suggest that two sessions (MSC09 session 07 and MSC10 session 02) may be outliers, as their distances to nearly all other sessions are relatively large across both distance metrics and all networks. Upon inspection, these sessions showed high levels of head motion. The mean framewise displacement (FD) for MSC09 session07 was 0.32 mm, with 70.9% of time points exceeding 0.2 mm, the threshold used in the preprocessing pipeline to define motion-contaminated time points (Gordon et al., 2017); for MSC10 session02, the mean FD was 0.39 mm, with 72.5% of time points exceeding the threshold. For all ROIs, AIRM tended to identify more pairs of sessions with large distances than the Frobenius metric did, as reflected by the darker bands in the matrix. Distance matrices based on Frobenius distance showed similar patterns across networks. These findings suggest that distance matrices may be useful for visualizing and identifying anomalous sessions.

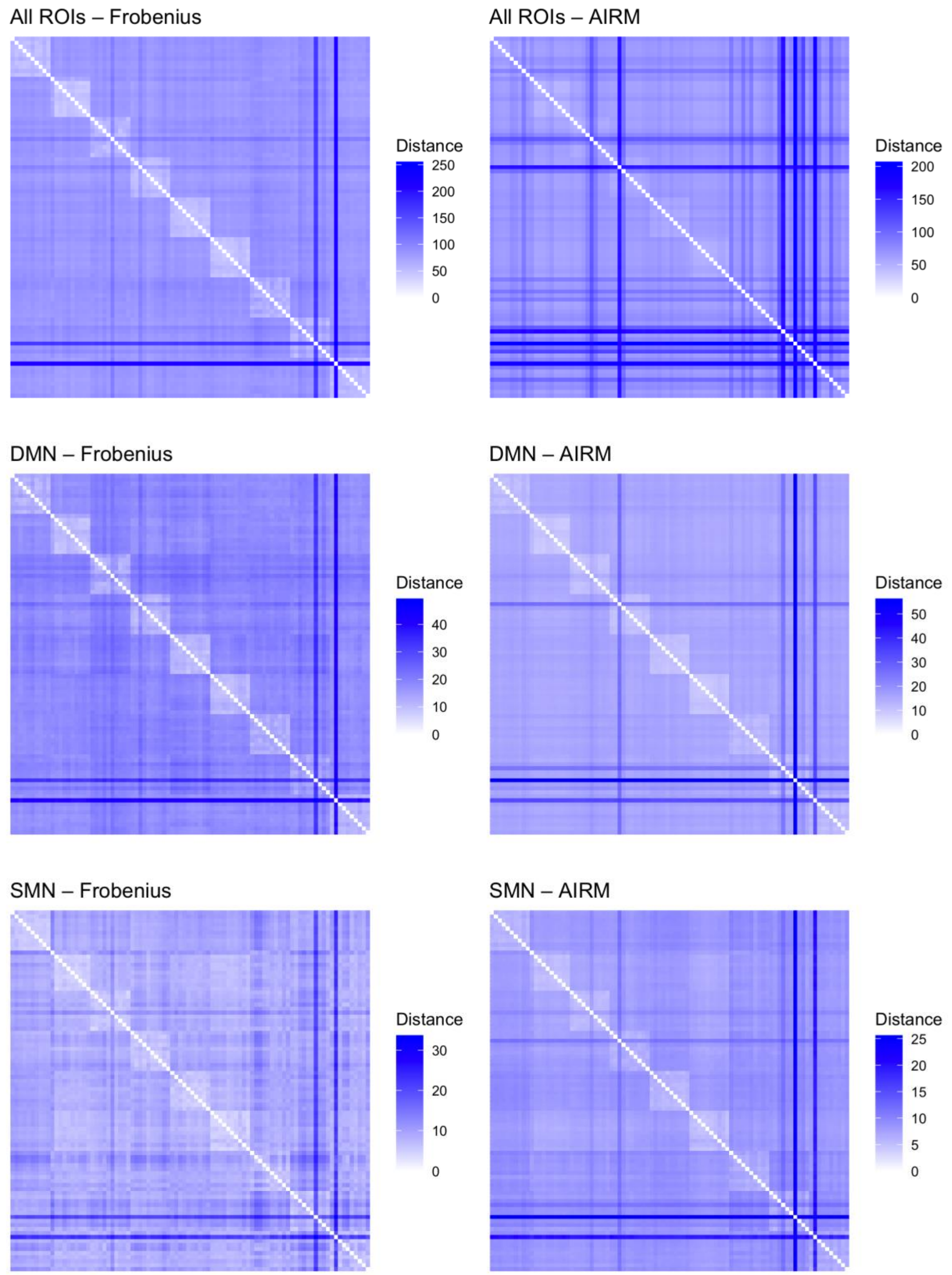

**Figure 1.** Distance matrices for all ROIs and two representative networks computed using two different metrics. Each row and column corresponds to a scan session; each cell represents the distance between a pair of sessions. Darker colors represent larger distance.

Distances between sessions from the same subject were generally smaller than those between sessions from different subjects, as shown by the nine lighter 10 × 10 squares along the diagonal. However, this within-subject structure was less apparent in the AIRM matrix for all ROIs.

### 3.2 Comparing Metrics

To further examine the differences between the two metrics, we visualized the relationship between pairwise distances computed using Frobenius metric and AIRM. **Figure 2** shows kernel density estimates of the within-subject and between-subject distance distributions under the Frobenius metric and AIRM. For All ROIs, the Frobenius metric demonstrated a clearer separation than AIRM between within-subject and between-subject distance distributions. In the DMN, both metrics yielded well-separated distributions. Similar patterns were observed in the FPN, CON, and VIS, as presented in **Supplementary Figure 2**. However, in the SMN, both Frobenius distance and AIRM showed considerable overlap between these two distributions.

**Figure 3** shows a scatterplot of distances calculated using each of the two metrics for the same session pairs. Each point represents a session pair with the Frobenius distance on the x-axis and the AIRM distance on the y-axis, colored by subject. Triangles indicate within-subject distances, while grey circles denote between-subject distances.

Overall, AIRM values tend to be positively correlated with Frobenius distance values. The pairwise distance scatterplot reveals several clusters: one larger cluster in the lower-left region and several smaller clusters in the upper region. Within each cluster, within-subject distances (colored triangles) tend to be smaller than between-subject distances (grey circles), with the within-subject points generally located toward the lower-left of each cluster. Notably, in each case the between-subject points lie slightly higher but more distinctly to the right of the within-subject points. This indicates that Frobenius distances provide stronger separation between within- and between-subject pairs.

Additionally, several distant clusters (primarily from MSC04, MSC09, and MSC10) were identified, consistent with the outlier sessions observed in the distance matrices. Furthermore, a separate group of points (from MSC04 session 03 and MSC09 session 04) appears relatively far from the main group according to the AIRM, though they would not be regarded as unusual according to the Frobenius distance.

When examining functional networks (**Figure 4** and **Supplementary Figure 3**), the positive correlation between AIRM and Frobenius distances appears more pronounced, while the relative positions of the main cluster and outlier clusters shift.

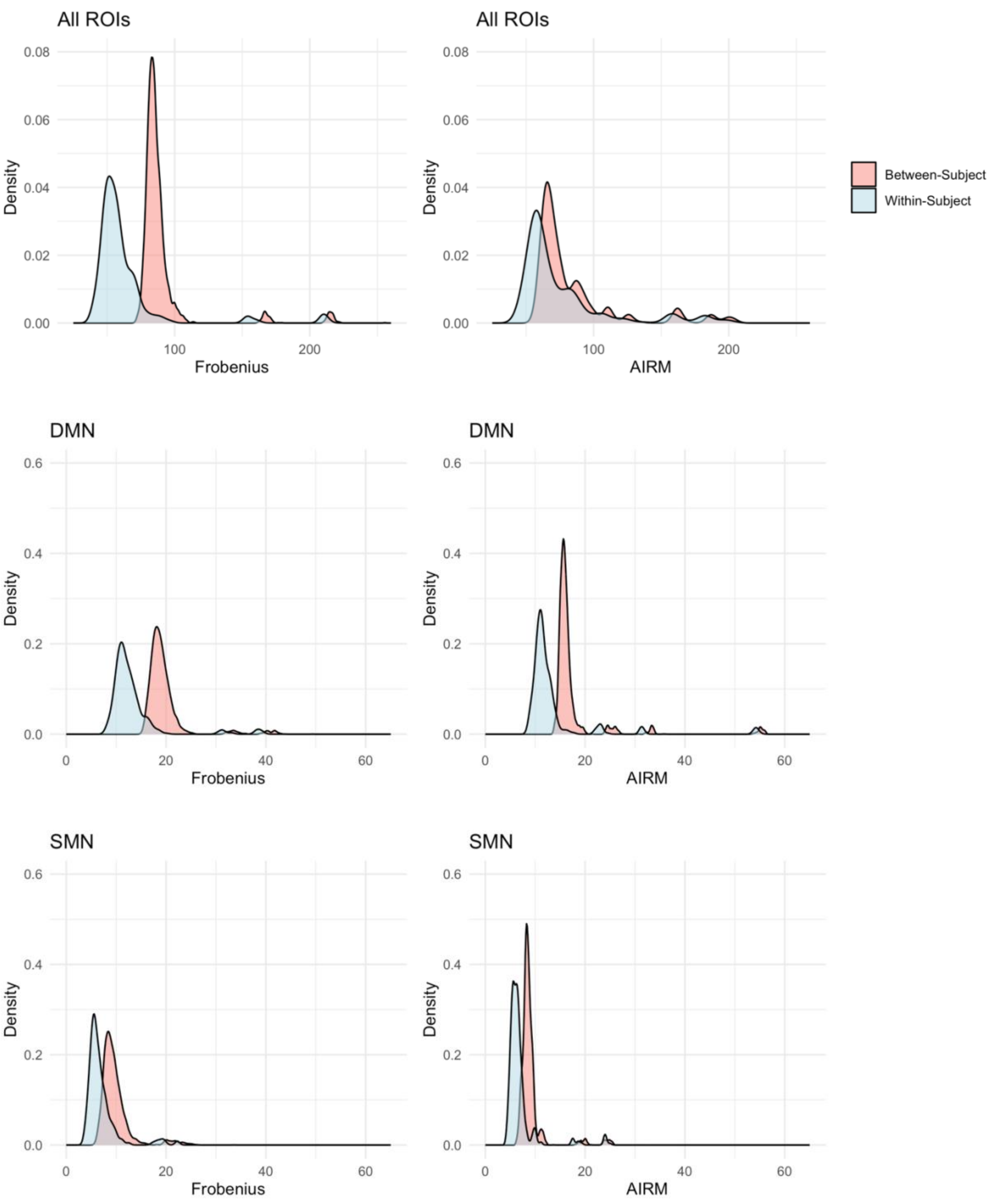

**Figure 2.** Density plots for all ROIs and two representative networks comparing between-subject (red) and within-subject distances (blue) using Frobenius metric and AIRM.

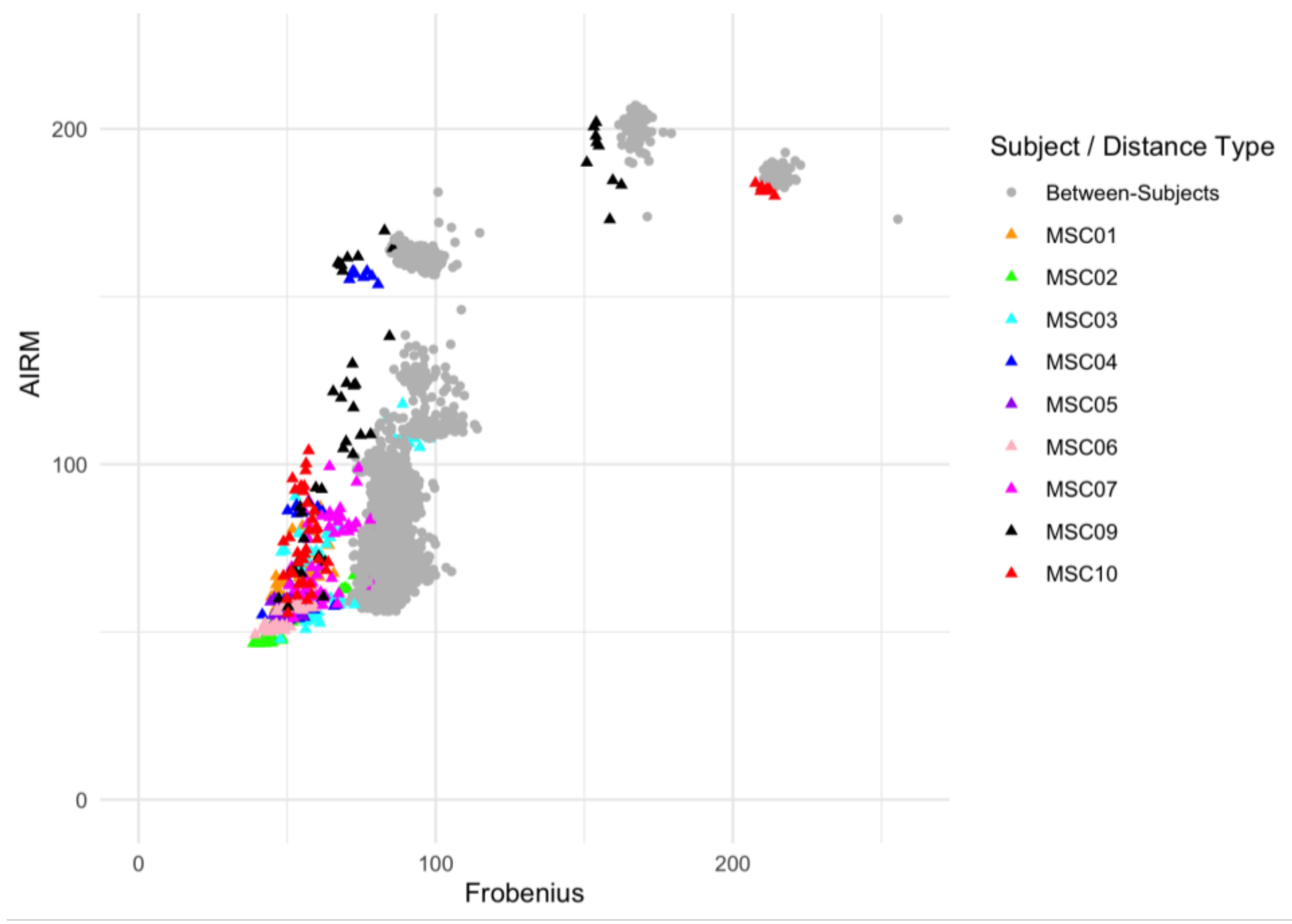

**Figure 3.** Scatterplot for all ROIs comparing pairwise distances calculated using Frobenius metric and AIRM. Each point represents distance between a pair of sessions. Triangles indicate within-subject distances, colored by subject ID, while grey circles represent between-subject distances.

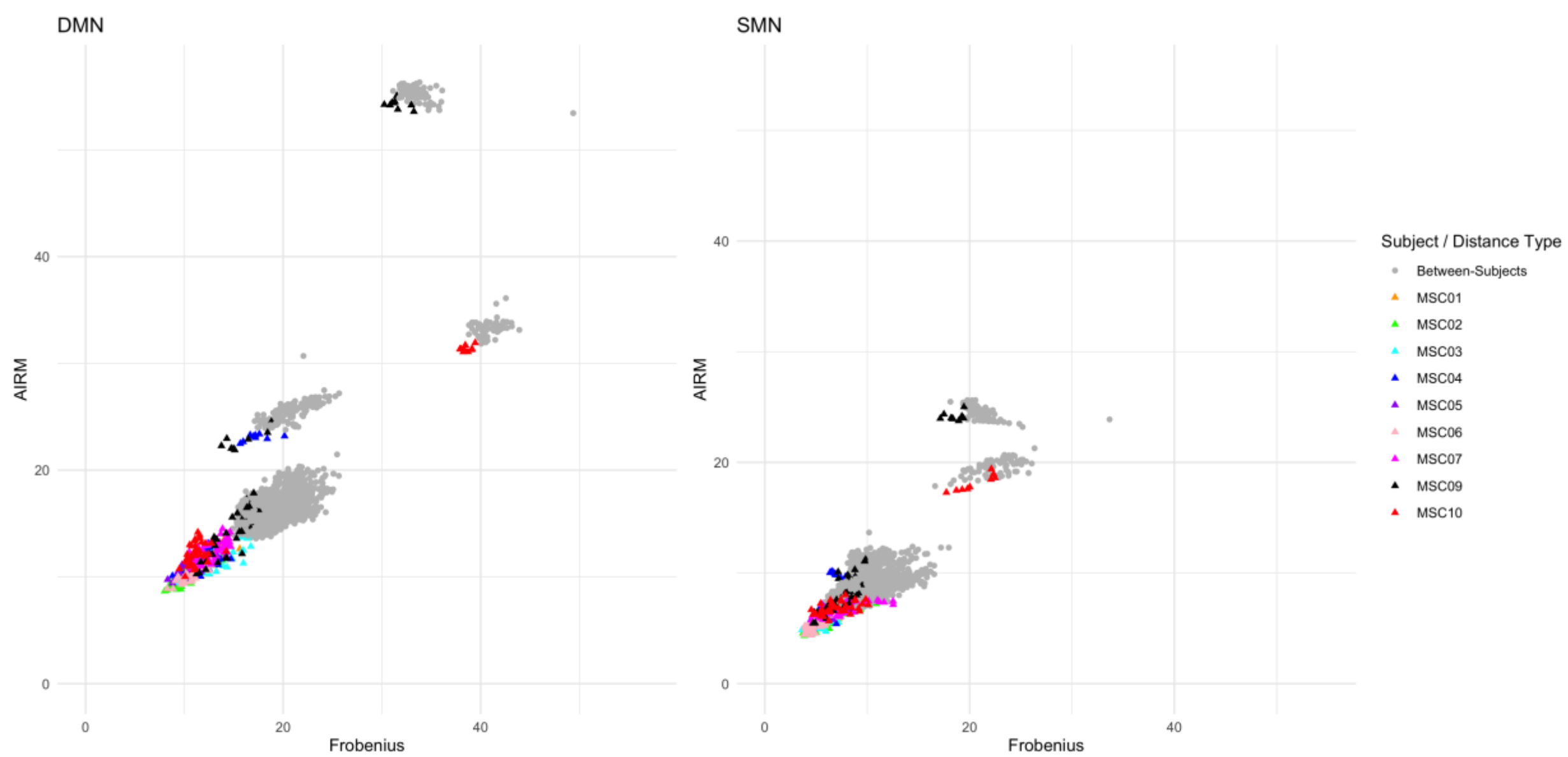

**Figure 4.** Scatterplot for two representative networks comparing pairwise distances calculated using Frobenius metric and AIRM.

### 3.3 dbICC in Different Networks

The dbICC values (**Table 1**) reflect the percentage of mean squared distances that is attributable to between-subject differences. Both distance metrics showed that the FPN had the highest dbICC. However, there was disagreement in SMN: under Frobenius metric, its dbICC was lower than that of All ROIs, whereas under AIRM, it ranked in the middle range among all networks.

To further understand these patterns, we examined the within-subject mean squared distance ($MSD_w$) and the between-subject mean squared distance ($MSD_b$) (**Table 1**). In functional networks, the $MSD_w$ were comparable between the Frobenius metric and AIRM. In contrast, larger discrepancies were observed for All ROIs and for $MSD_b$ across functional networks.

**Table 1.** dbICC, bootstrap confidence intervals, $MSD_w$, and $MSD_b$ computed under Frobenius metric and AIRM, across all ROIs and functional networks.

| Networks | Number of ROIs | Frobenius | | | AIRM | | |
|---|---|---|---|---|---|---|---|
| | | dbICC (95% CI) | $MSD_w$ | $MSD_b$ | dbICC (95% CI) | $MSD_w$ | $MSD_b$ |
| All ROIs | 360 | 0.454 (0.340, 0.573) | 4718.46 | 8638.99 | 0.156 (0.074, 0.238) | 7006.96 | 8298.04 |
| DMN | 77 | 0.502 (0.390, 0.611) | 199.94 | 401.61 | 0.342 (0.179, 0.486) | 232.23 | 352.68 |
| FPN | 50 | 0.592 (0.464, 0.719) | 84.88 | 208.20 | 0.489 (0.332, 0.625) | 90.083 | 176.22 |
| CON | 56 | 0.576 (0.465, 0.692) | 104.26 | 246.09 | 0.379 (0.214, 0.526) | 109.55 | 176.46 |
| VIS | 60 | 0.575 (0.482, 0.664) | 128.27 | 301.83 | 0.393 (0.256, 0.508) | 129.57 | 213.61 |
| SMN | 39 | 0.450 (0.333, 0.549) | 58.15 | 105.73 | 0.357 (0.229, 0.483) | 59.46 | 92.47 |

### 3.4 Effect of Scan Length on dbICC

**Figure 6** shows how dbICC values change with increasing scan length. For both distance metrics and across all networks, dbICC values increased as scan length increased, indicating that longer scans yield more reliable estimates. The most pronounced improvement occurred within the first 10–20 minutes, after which the curves tended to plateau. **Figure 7** shows an approximately linear relationship between $log\ (m)$ and $log(\frac{\hat{\rho}_m}{1-\hat{\rho}_m})$. Slopes vary by metric and network, with greater variability observed under AIRM. Detailed slope estimates are provided in **Supplementary Table 1**.

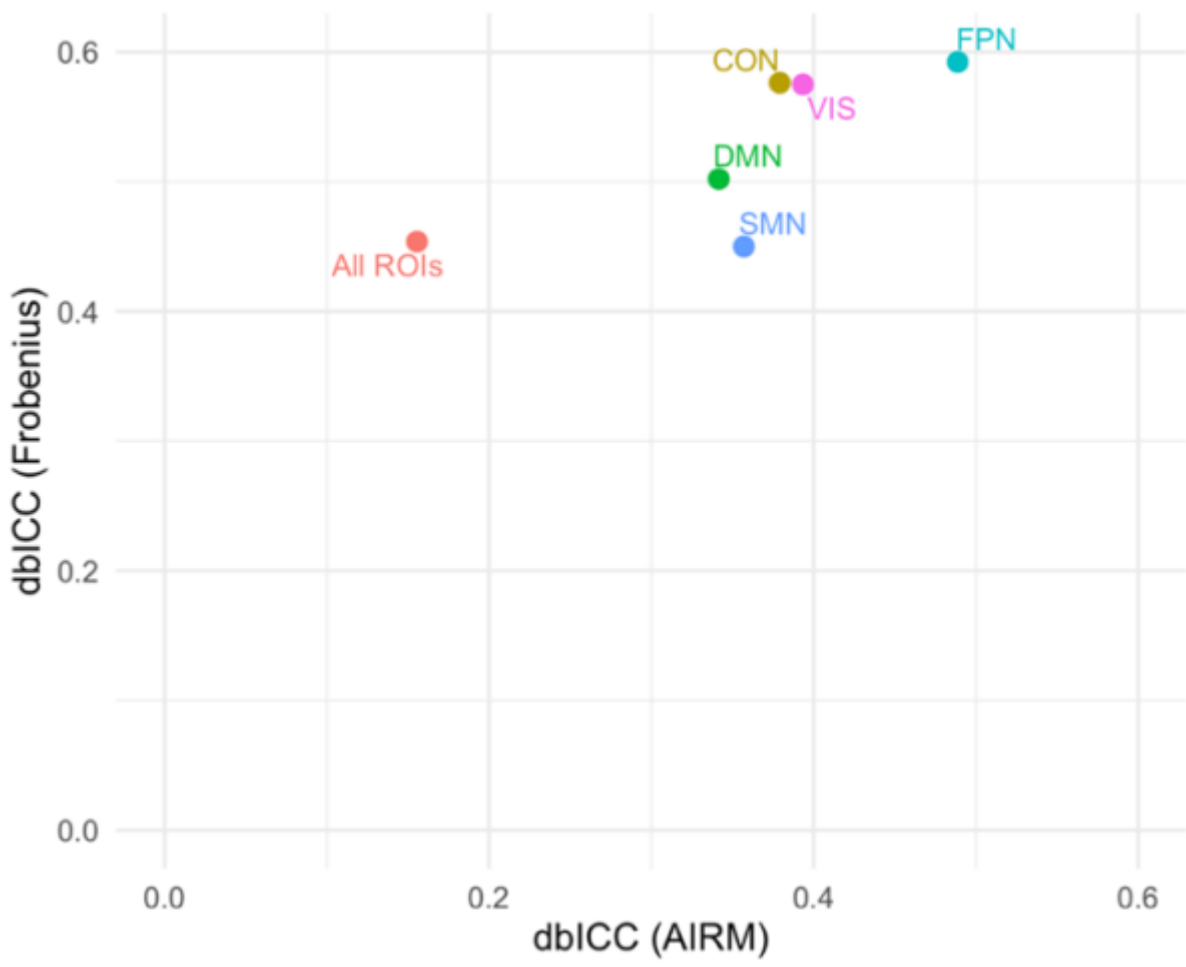

**Figure 5**. dbICC values under Frobenius metric and AIRM across different networks.

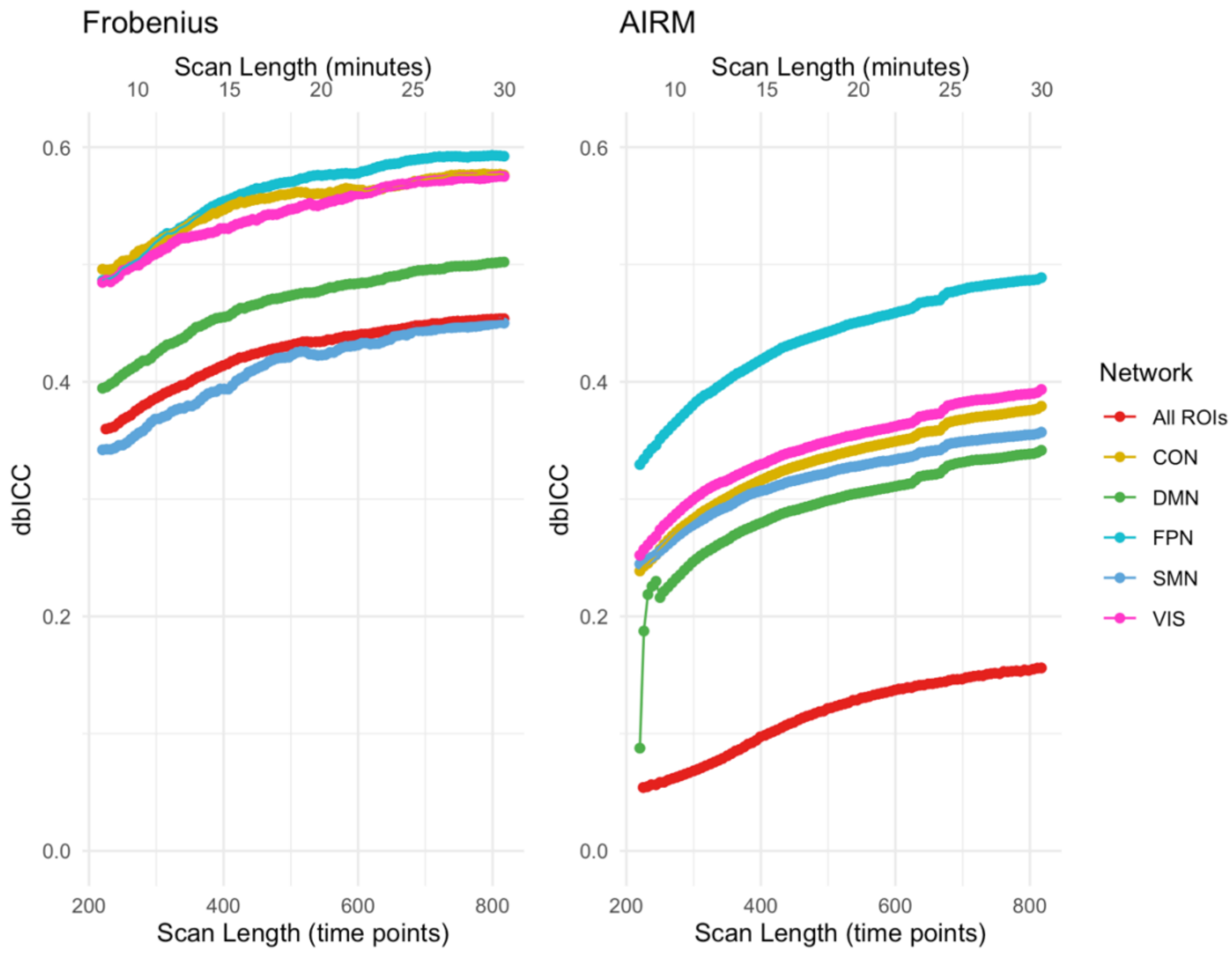

**Figure 6.** Scan length and dbICC values under Frobenius metric and AIRM, across all ROIs and functional networks. Each curve represents the dbICC values computed at different scan lengths for a specific network. The bottom x-axis is accumulated time point (TR = 2.2s) and the top x-axis is the scan length in minutes.

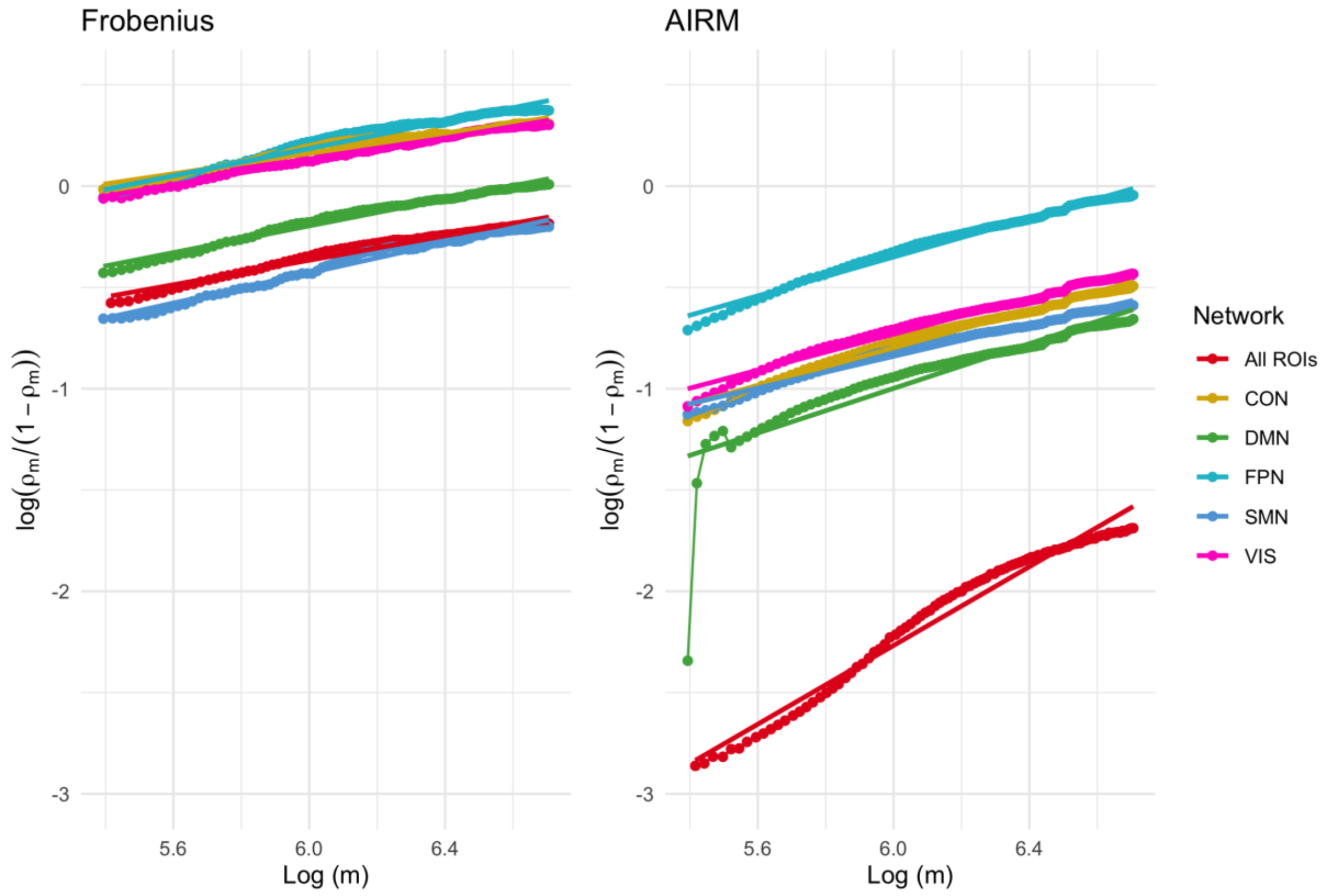

**Figure 7.** Log-log plot for scan length and dbICC values under Frobenius metric and AIRM, across all ROIs and functional networks. The straight lines are the fitted linear trends from regressing $log(\frac{\hat{\rho}_m}{1-\hat{\rho}_m})$ on $log\ (m)$ for each network.

### 3.5 Effect of Time Interval on dbICC

To reduce the influence of outlier sessions, we excluded MSC09 session 07 and MSC10 session 2, which were identified as apparent outliers in the distance matrices according to both metrics. From the result of model (3), we observed significant associations between time interval and squared within-subject distance in all ROIs ($p = 0.048$), CON ($p = 0.050$), VIS ($p = 0.006$) and SMN ($p = 0.023$) only under the Frobenius distance metric. No other time interval effects were significant at the 0.05 level in other networks or when using AIRM. **Supplementary Table 2** reports the $p$-values for all sessions and after excluding the two outlier sessions, and **Supplementary Table 3** summarizes dbICC estimates by inter-session interval.

## 4 Discussion

### 4.1 Measured Reliability Depends on the Choice of Metric

In this study we have found that the reliability of rs-fMRI functional connectivity, as measured by dbICC, is sensitive to the choice of metric. Across all networks, Frobenius metric generally yielded higher dbICC values. For functional networks (DMN, FPN, CON, VIS, and SMN), the discrepancy between metrics tended to be less pronounced, possibly due to reduced complexity in the correlation structure and lower impact of noise.

This discrepancy may reflect how each metric handles the structure of correlation matrices. AIRM respects the Riemannian geometry of SPD matrices, while Frobenius metric ignores it. As a result, the dbICC values may emphasize different aspects of variability, with AIRM reflecting intrinsic geometric relationships and Frobenius reflecting Euclidean differences. In addition to geometric considerations, the presence of noise may also influence reliability estimates. Additive noise arising from time-series measurements in fMRI is naturally placed within Euclidean space, and so in such a case the Frobenius metric may exhibit greater robustness. In contrast, AIRM may be more sensitive to perturbations that alter the geometric structure, which can affect the stability of the estimates.

Considering the manifold structure of the space of connectivity matrices, geometry-aware metrics such as AIRM may be appropriate for measuring distances between connectivity matrices. However, AIRM is relatively computationally expensive, which may limit its practical use. Moreover, AIRM is not the only Riemannian metric on SPD matrices; for example, the log-Euclidean metric provides a computationally simpler alternative, although it does not have the same affine-invariance property. Thus, the optimal choice of metric depends on both analysis goals and computational feasibility, as well as the level and structure of noise in the data. Future studies could assess the relative performance of the various metrics in classification problems and investigate how different noise levels and structures influence their behavior, which may help understand their relative strengths in different settings.

A further geometric consideration concerns the specific structure of correlation matrices. While AIRM is designed for the space of SPD matrices, correlation matrices additionally carry the constraint that all diagonal entries equal one, which restricts them to a subset of the SPD manifold known as the elliptope (Laurent & Poljak, 1995). The geometry of the elliptope differs from that of the full SPD manifold, and metrics defined on SPD matrices are not necessarily optimal within this more constrained space. Thanwerdas (2024) proposed metrics intrinsic to the elliptope, including the log-scaled metric, which is specifically designed to respect the geometric structure of full-rank correlation matrices. We conducted preliminary investigations in this direction; however, the results were not stable across settings and proved difficult to interpret, likely due to the sensitivity of these metrics to noise, complex boundary behavior of the elliptope, and the additional complexity of their estimation. We therefore leave a systematic exploration of elliptope-aware metrics for dbICC estimation to future research.

### 4.2 Functional Networks have higher dbICC compared to Larger Networks

At the network level, functional networks tended to show higher reliability. This may be due to greater functional homogeneity within these networks and the reduced influence of noise and irrelevant connections compared to whole-brain analyses (Jiang & Zuo, 2016), which involve estimating over 64,000 pairwise connections. Additionally, we observed that some correlation matrices, particularly in the whole-brain condition, were not positive definite, likely due to unstable estimation of the correlation matrices caused by an insufficient number of time points relative to the large number of ROIs (Liégeois et al., 2020). Therefore, regularization was required prior to computing AIRM distances. While technically necessary, this step may influence results and needs further evaluation.

The SMN showed unique characteristics in our analysis. Specifically, it exhibited highly consistent connectivity patterns not only within subjects but also across different subjects, leading to small $MSD_w$ and $MSD_b$. Thus, the low dbICC reflects low between-subject variability rather than high within-subject variability.

### 4.3 dbICC Changes with Scan Length and Time Interval

Regarding scan length, we observed that dbICC increased with longer durations, especially during the first 20 minutes. This is consistent with prior findings (Birn et al., 2013; Noble et al., 2017) and suggests that 20 minutes may be an optimal scan length, offering a good balance between reliability, scanning cost, and participant burden.

We observed significant associations between time interval and within-subject distance in the VIS and SMN under the Frobenius metric. However, no such effects were found in other networks or under the AIRM. Given the exploratory nature of the analysis, the lack of correction for multiple testing, and the potential non-independence of observations, these results should be viewed as preliminary. Overall, we found no consistent evidence that longer time intervals reduce reliability in rs-fMRI over the relatively short intervals examined in this study. This finding suggests that functional connectivity is relatively stable across sessions separated by up to approximately one week. However, our data do not address whether reliability would decrease over longer intervals, such as several weeks or months. In addition, when interpreting these results we should also take into account that correlation analysis typically relies on the assumption of temporal stationarity (Liégeois et al., 2017), i.e., that each subject's brain functional organization remains the same over time. Future studies involving longer intervals are needed to further test this assumption.

The MSC dataset provided a valuable opportunity to examine the effect of scan length, as it offers relatively long 30-minute resting-state scans across multiple sessions. However, the small number of subjects (n = 9) remains a limitation, which may affect the stability and generalizability of the scan length and time interval estimates. For example, the distribution of time intervals was uneven, with fewer session pairs at longer gaps (e.g., five to seven days), which may affect the precision of estimates for those ranges.

## 5. Conclusion

This study evaluated the reliability of rs-fMRI functional connectivity using dbICC under different metrics and scanning conditions, based on data from the MSC dataset. We found that dbICC estimates vary with the choice of distance metric. dbICC values improved with longer scan lengths—particularly within the first 20 minutes—while the effect of time intervals was minimal.

## Ethics statement

This study used publicly available, de-identified data from the Midnight Scanning Club (MSC) dataset. Ethical approval was not required. Large language models such as ChatGPT were used to correct grammar mistakes and improve clarity.

## Data and code availability

The MSC dataset can be accessed at https://openneuro.org/datasets/ds000224/versions/1.0.4. Analysis code is available from the corresponding author upon request.

## Author Contributions

**Yu Huang** conducted the data analysis, generated the figures and tables, and drafted the manuscript. **R. Todd Ogden** supervised the project, provided guidance throughout the study, and revised the manuscript. **Philip T. Reiss** contributed important ideas, provided methodological guidance, and revised the manuscript. **Seonjoo Lee** performed the initial data preprocessing, provided methodological guidance, and revised the manuscript.

## Declaration of competing interests

The authors have no competing interests to declare.

## Acknowledgements

We thank Yi Zhao from Indiana University School of Medicine and members of the Functional Data Analysis Working Group at Columbia University for their helpful feedback and discussions.

## Supplemental Material

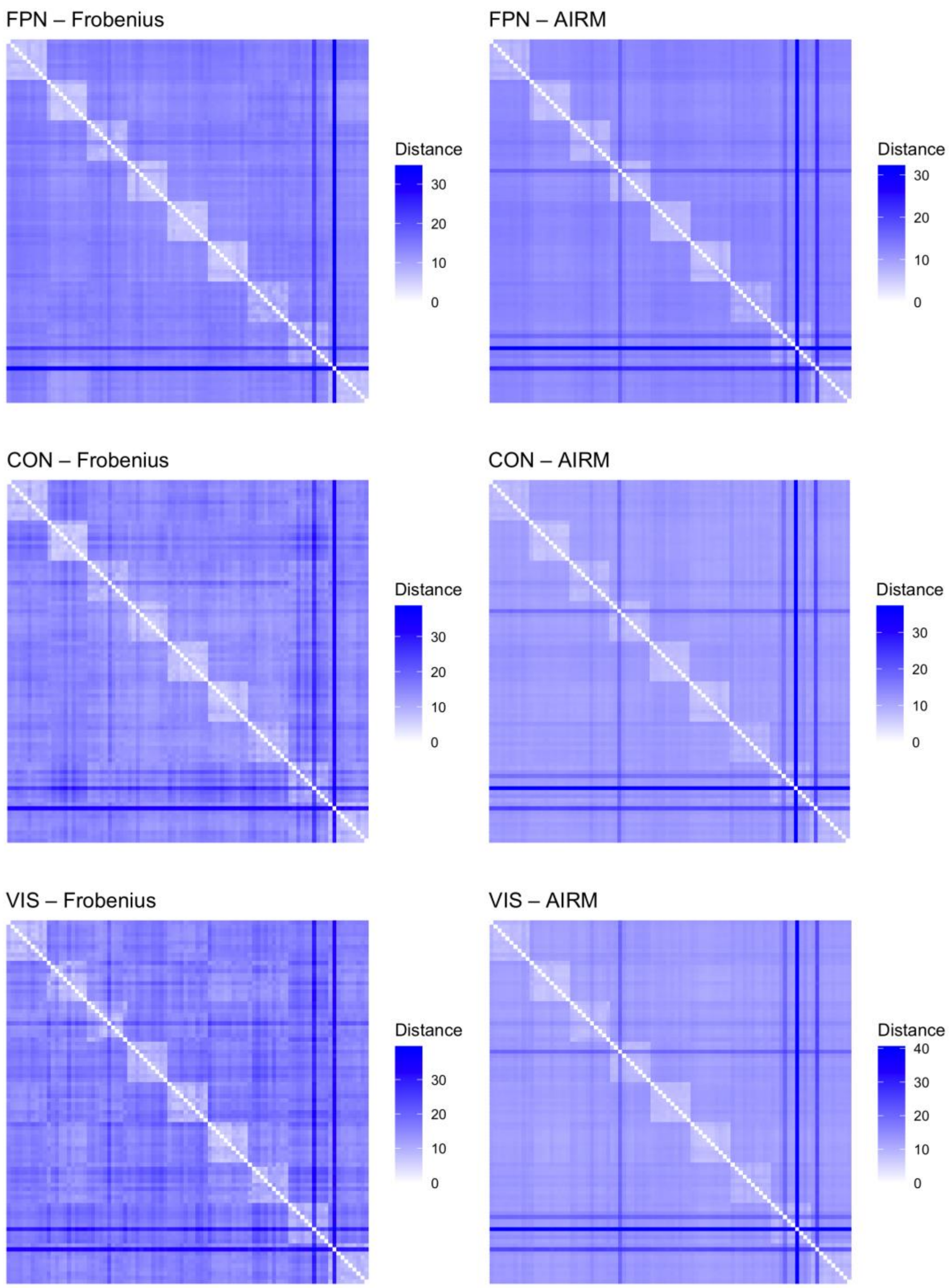

**Supplementary Figure 1.** Distance matrices for three networks computed using two different metrics. Each row and column corresponds to a scan session; each cell represents the distance between a pair of sessions. Lighter colors represent smaller distances, and darker colors represent larger distance.

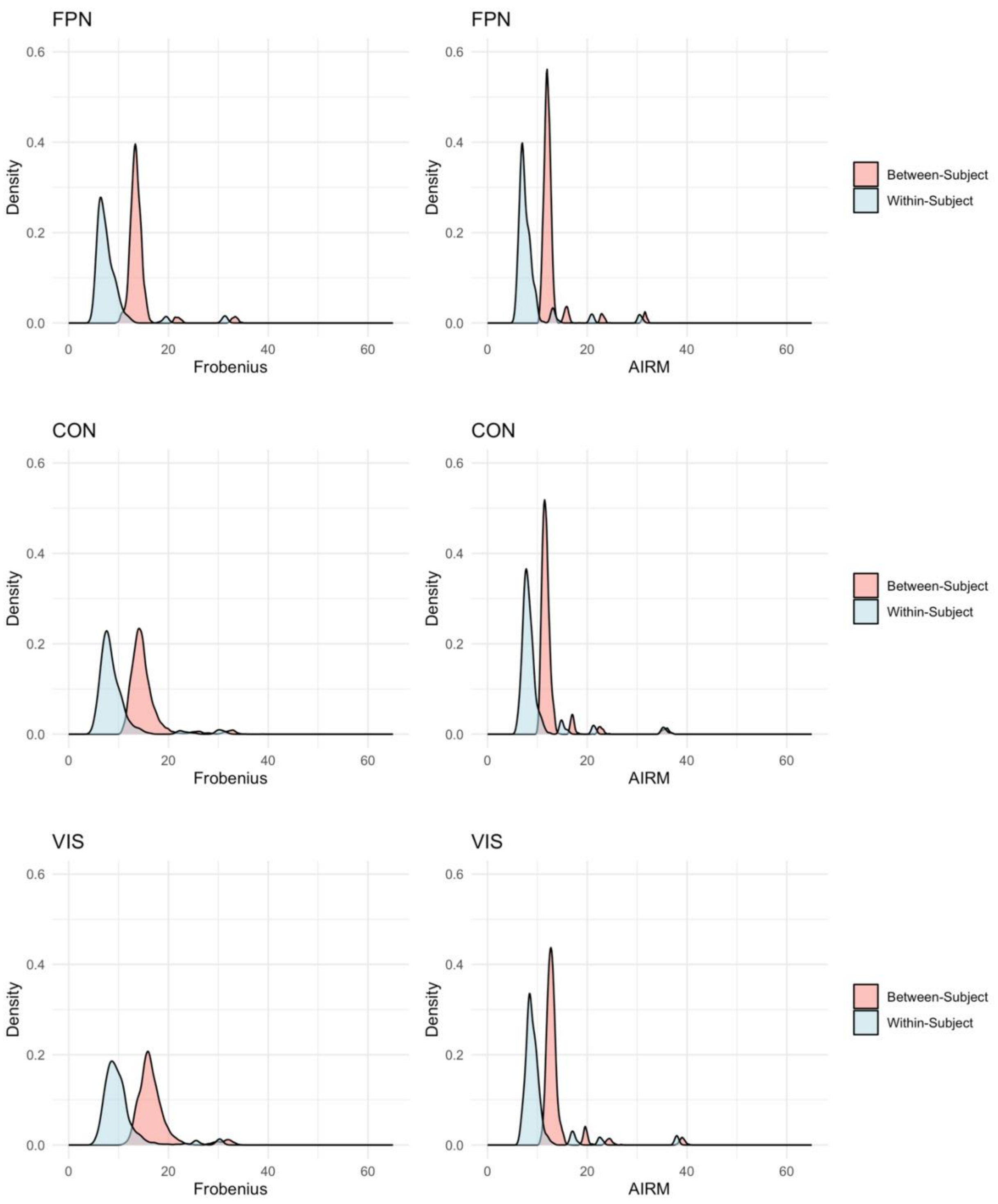

**Supplementary Figure 2.** Density plots for three networks comparing between-subject (red) and within-subject distances (blue) using Frobenius metric and AIRM.

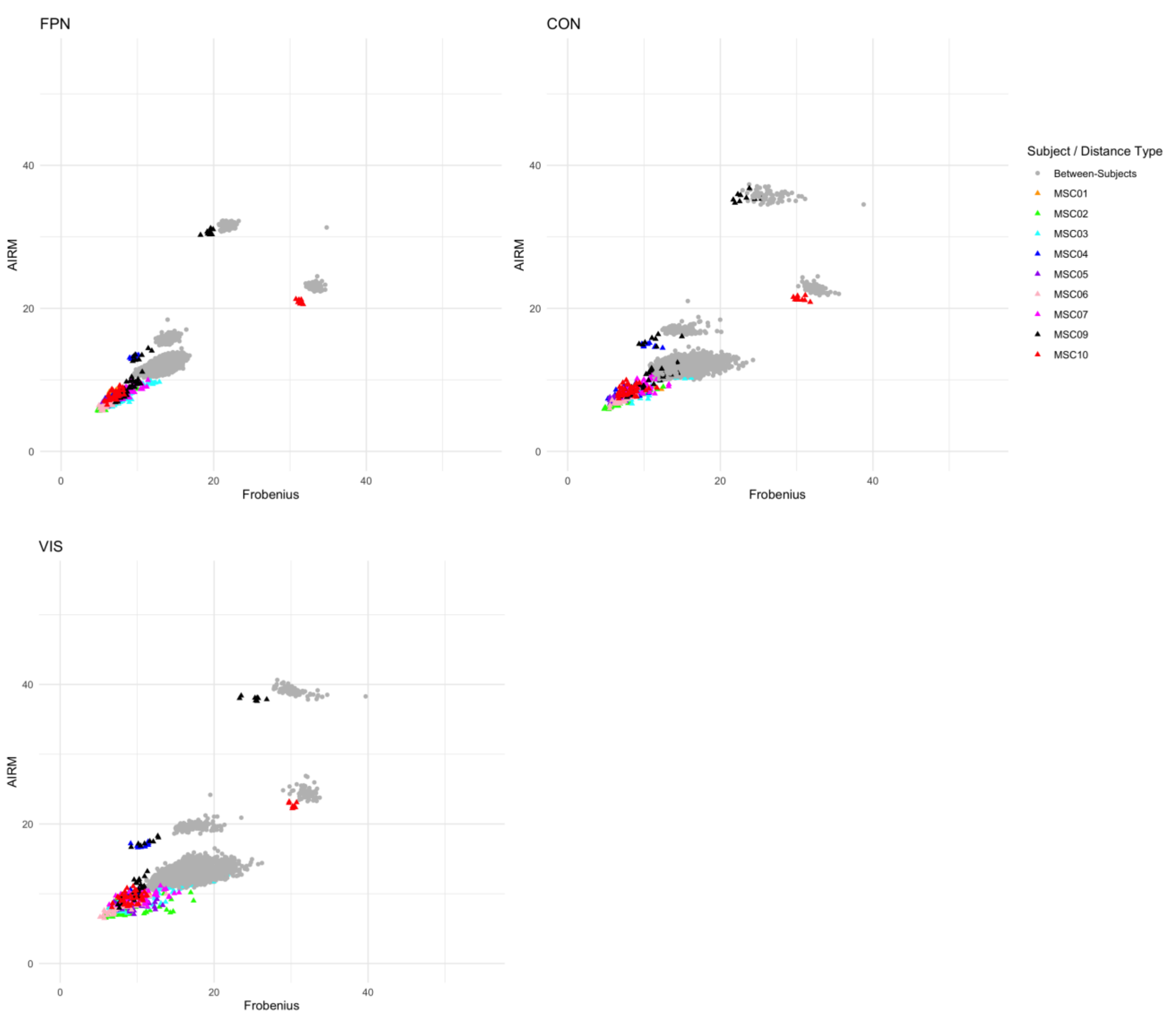

**Supplementary Figure 3.** Scatterplot for three networks comparing pairwise distances calculated using Frobenius metric and AIRM. Each point represents a distance between session pair. Triangles indicate within-subject distances, colored by subject ID, while grey circles represent between-subject distances.

**Supplementary Table 1**. Estimated slopes from log–log plots of reliability versus scan length across networks and metrics.

| | Frobenius | AIRM |
|---|---|---|
| All ROIs | 0.310 | 0.967 |
| DMN | 0.330 | 0.552 |
| FPN | 0.336 | 0.479 |
| CON | 0.250 | 0.464 |
| VIS | 0.286 | 0.445 |
| SMN | 0.371 | 0.391 |

**Supplementary Table 2**. P-values from GAM Model Assessing the Effect of Time Interval on Within-Subject Distance Across Distance Metrics and Networks, With and Without Outlier Sessions

| | All sessions | | Without two outlier sessions | |
|---|---|---|---|---|
| | Frobenius | AIRM | Frobenius | AIRM |
| All ROIs | 0.627 | 0.768 | 0.048 | 0.621 |
| DMN | 0.692 | 0.693 | 0.151 | 0.871 |
| FPN | 0.730 | 0.802 | 0.371 | 0.691 |
| CON | 0.597 | 0.732 | 0.050 | 0.832 |
| VIS | 0.324 | 0.795 | 0.006 | 0.530 |
| SMN | 0.367 | 0.959 | 0.023 | 0.285 |

**Supplementary Table 3.** dbICC Estimates Across Distance Metrics and Networks for 1-, 3-, 5-, and 7-day inter-session intervals

| Interval days | Frobenius | | | | AIRM | | | |
|---|---|---|---|---|---|---|---|---|
| | 1 | 3 | 5 | 7 | 1 | 3 | 5 | 7 |
| All ROIs | 0.559 | 0.543 | 0.526 | 0.509 | 0.121 | 0.138 | 0.154 | 0.170 |
| DMN | 0.583 | 0.573 | 0.562 | 0.552 | 0.433 | 0.430 | 0.427 | 0.425 |
| FPN | 0.687 | 0.681 | 0.676 | 0.671 | 0.578 | 0.574 | 0.570 | 0.567 |
| CON | 0.669 | 0.653 | 0.636 | 0.620 | 0.468 | 0.465 | 0.463 | 0.460 |
| VIS | 0.670 | 0.642 | 0.615 | 0.587 | 0.481 | 0.473 | 0.464 | 0.456 |
| SMN | 0.543 | 0.513 | 0.484 | 0.454 | 0.447 | 0.436 | 0.425 | 0.414 |